\documentclass{myart}

\usepackage[utf8]{inputenc}
\usepackage{graphicx}
\usepackage{pgfplots}
\usepackage{pgf-pie}
\usepackage{adjustbox}
\usepackage{textcomp}
\usepackage{pbalance}

\newcommand{\code}[1]{{\small\textsf{#1}}}

\newcommand{\hide}[1]{}

\title{The Paradox of Function Header Comments%
\thanks{This research was supported by the ISRAEL SCIENCE FOUNDATION (grant no.\ 832/18).}}

\author{Arthur Oxenhorn ~~~ Almog Mor ~~~ Uri Stern ~~~ Dror G. Feitelson\\[2mm]
\textit{Department of Computer Science}\\
\textit{The Hebrew University, Jerusalem, Israel}
}

\begin{document}
\widowpenalty=0 \displaywidowpenalty=100 \clubpenalty=0
\predisplaypenalty=100 \postdisplaypenalty=100

\maketitle

\begin{abstract}
    Perhaps the most widely used form of code documentation is function header comments.
    We performed a large-scale survey of 367 developers to catalog their expectations from such documentation and to chronicle actual practice.
    Paradoxically, we found that developers appreciate the value of header comments and estimate that they are worth the investment in time, but nevertheless they tend not to write such documentation in their own code.
    Reasons for not writing header comments vary from the belief that code should be self-documenting to concern that documentation will not be kept up-to-date.

    A possible outcome of this situation is that developers may evade requirements to write documentation by using templates to generate worthless comments that do not provide any real information.
    We define a simple metric for information-less documentation based on its similarity to the function signature.
    Applying this to 21,140 files in GitHub Python projects shows that most functions are undocumented, but when header comments \emph{are} written they typically do contain additional information beyond the function signature.\\[2mm]
\end{abstract}

\section{Introduction}

When faced with the task of understanding a computer program, the two chief sources of information are the program's code itself and the documentation of this code.
The documentation in the code can be divided into two main parts: class and function header comments, and inline comments \cite{jiang06}.
Function header comments in particular are used to describe the algorithm implemented by the function or to document its interface, in both cases with the goal to eliminate or at least reduce the need to read and comprehend the code itself.
Thus they support a top-down approach to code comprehension \cite{woodfield81,arab92,parnas94,meng19}.
Inline comments complement this by explaining elements of the code that may not be obvious on their own right.
This is especially useful when trying to comprehend the actual code, for example in the context of maintaining it.

The potential importance of comments is generally recognized, and writing header comments is often required in coding guidelines.
On the other hand, many have also expressed concern regarding the quality and relevance of comments \cite{steidl13,uddin15,head18,wang23}.
Berry et al.\ note the problem that documentation is usually beneficial for its reader, not its writer, thereby reducing the incentive to write good comments \cite{berry16}.
A related issue is the need to update comments as code evolves \cite{lethbridge03}; a possible solution is to only produce the comments at the code review stage, when the code is being finalized \cite{head18}.
Other studies claim that documentation just does not have a significant effect on code comprehension \cite{borstler16}.

Functions are the most-often commented code elements \cite{fluri07}.
The practice of function documentation depends on the programming language used.
In C and C++ a distinction is made between comments in header files, which document the interface, and comments in code files, which document the implementation.
In Java it is common to format header comments for JavaDoc, which automatically generates external documentation of APIs \cite{kramer99}.
Python and other languages support docstrings, that remain available also during execution.
These practices are intended to support developers who need to use functions written by others.

But in the last 20 years or so the notion that code should be self-documenting has gained traction.
This approach to programming emphasizes the need to use clear variable and function names, with the aim of not having to add any comments at all.
Steve McConnell, in a whole chapter devoted to writing functions in \textit{Code Complete}, does not even mention header comments \cite{mcconnell:cc}.
He does mention them elsewhere as a first step in writing pseudocode leading to an implementation.
As for using documentation in general, McConnell strongly opposes predefined enforced header comment guidelines, preferring that documentation reflect the complexity of the function.
Parameters should only be described if one is using an automatic documentation-generation facility such as JavaDoc.
Robert Martin, in \textit{Clean Code} --- which also has a whole chapter on functions --- is more terse.
His subsection on header comments contains only two sentences:
``Short functions don’t need much description. A well-chosen name for a small function that does one thing is usually better than a comment header.''

Given such widely diverging views on the utility and need of header comments, we turn to see what developers do in practice, and how they perceive header comments in their everyday work.
What should function header comments, and API documentation in general, contain?
What do developers need and want?
And do they in fact act according to what they say they want?

The contributions of this paper are the following:
\begin{enumerate}
    \item We conduct a large-scale survey of developers, and collect information on their practices concerning the use and production of function header documentation, what they find most useful in such documentation, and what is most often missing;
    \item We uncover paradoxical behavior, where developers tend to use and value function header comments and think they are worth the investment, but nevertheless tend not to write them;
    \item We refine the definition of a simple metric for determining whether function header documentation contains additional information beyond the function signature;
    \item We collect data about the use of function header comments in Python projects, showing that most functions are undocumented, but in most documented functions the documentation does contain novel information beyond the function signature.
\end{enumerate}

\section{Related Work}

\hide{check
Pooja Rani, Arianna Blasi, Nataliia Stulova, Sebastiano Panichella, Alessandra Gorla, Oscar Nierstrasz 
A decade of code comment quality assessment: A systematic literature review
Journal of Systems and Software, Volume 195, September 2022, Article Number 111515
X Hu, G Li, X Xia, D Lo, Z Jin
Deep code comment generation
Proceedings of the 26th conference on program comprehension, 200-210
}
Documentation is widely accepted as an element of programming, and there has been a significant amount of work on how to write comments and on their effect.
Steidl at al.\ present a classification of comment types and suggest simple metrics for their quality \cite{steidl13}.
Aggarwal et al.\ base their model of software maintainability on three aspects of documentation: the percentage of comment lines, the simplicity of the comments text, and the correlation between documentation vocabulary and the code vocabulary \cite{aggarwal02}.
Other research indicates that documentation complements the information contained in variable names \cite{blinman05}, and that it can compensate for low code quality \cite{lemos20}.
Even redundant documentation (for example explicitly stating what design pattern is being used) may still help somewhat \cite{prechelt02}.
Documentation's utility may also depend on experience: novices appear to focus on comments more than experienced programmers \cite{crosby02}.

Function header comments used to document APIs (Application Programming Interfaces) have received special attention.
Liskov and Guttag suggest that functions need a specification comment containing a pair of assertions: the function's precondition and its postcondition \cite{liskov:book}.
This is part of the abstraction (and specifically procedural abstraction) that also supports formal verification and testing, and is thought to be sufficiently informative to allow others to use the function without looking at its body.
Meyer makes a similar suggestion, in the form of function contracts \cite{meyer92}.
In addition to documentation, the function name itself if considered important for understanding what a function does.
Alsuhaibani el al.\ conducted a large survey of developers, and found that there is large agreement on the basic principles of function naming \cite{alsuhaibani21}.

Other research has looked at how developers use API documentation.
This is usually based on surveys of developers, similar in spirit to our work \cite{head18,meng19}.
Watson et al.\ complement surveys with collecting data from actual practice in open source projects \cite{watson13}.

Research that uncovered problems in API documentation can be especially revealing.
Robillard found that problematic or missing documentation is only one of the problems in learning to use an API \cite{robillard09}.
Other problems include the API itself (e.g.\ if it is badly designed) and lack of background knowledge.
Moreover, deficiencies in API documentation may cause the API to be abandoned and another to be used in its place \cite{uddin15}.
Watson et al.\ add that having required elements is not enough for effective documentation: issues like the writing style, visual design, and ease of finding the documentation are also important \cite{watson13}.
Curtis et al.\ found that using a restricted language forces consistency that can be helpful \cite{curtis89}.

Finally, in a survey of research on documentation, Rai et al.\ note that a favorite topic is the generation of summaries of functions \cite{rai22}.
For example, Huang et al.\ describe a method that combines information retrieval with neural networks for the purpose of generating a general description of a function \cite{huang22}.
More recently large language models are also being applied to this problem.

\section{Research Questions}

The general goal of our research was to learn whether function header documentation indeed helps developers in understanding functions' usage, thereby saving the effort to read and understand the actual code.
If it turns out that this is \emph{not} the case, then the effort invested in writing comments has been wasted, and doing so can be avoided.

To gain insights into this issue, we formulated the following concrete research questions:
\begin{enumerate}
    \item To what degree is function header documentation used in practice?
    What fraction of functions have such documentation?
    What fraction of developers rely on such documentation, and what fraction actually invest in writing header comments in their code?
    These questions will show whether this topic is important at all, and specifically how developers see it.
    \item Which parts of function header documentation (parameters, return value, examples) are the most useful?
    How often are they actually included in header comments?
    This may shed light on when developers can make do with the documentation, and when they also need to read the code.
    \item What necessary information is most often missing from function header documentation, thereby making it ineffective?
    Answering this question can provide guidance concerning how header comments can be improved.
    \item In general, does function header documentation add meaningful information beyond the function's signature?
    Or is it redundant?
    This will tell us whether function and parameter names are meaningful enough to eliminate the need for header comments.
\end{enumerate}

To answer these questions we employed two methodologies: a survey and repository mining.
The survey was answered by 367 developers from around the world.
It had two main parts.
In one we asked about perceptions regarding header comments: how useful they were thought to be, and which parts (e.g.\ the explanation of parameters, return value, or internal structure) were more important than others.
In the other we asked about actual practice: whether functions in the project they work on have header comments, whether they themselves write such comments, and why.

In the repository mining we scanned the code in 21,140 files from 68 top-ranked python projects hosted on GitHub.
We extracted the doc strings of the functions in these files, and compared them with the function signatures.
This provides a simple measure of whether the documentation contains any additional information beyond the function signature itself.

\section{Survey on Perception and Use of Header Comments}

We conducted a survey among developers from different backgrounds and locations, to see what they think of header comments and how they use them.

\subsection{Survey Design}

The survey was composed of the following sections:
\begin{enumerate}
    \item Questions on how much developers read and rely on function header documentation.
    This part included questions on whether they read this documentation first; if so, what do they most focus upon (explanation of the parameters, the function's inner methods, its return value, or usage examples), and does reading the documentation suffice to understand the function's usage; if not, why not, and whether they return to read it at a later stage.
    \item Questions on how useful developers find the information in header comments to be (even if they do not actually read it for various reasons), and which parts of it are the most useful ones.
    An additional question concerned what information is usually lacking from such documentation.
    \item Questions about how much effort developers spend in writing and updating function header comments, and whether they feel it is worth the time this consumes.
    \item Questions on the current practice of documenting code in their workplace.
    For this we asked about the existence of company and team guidelines for documenting code.
    \item Lastly, we asked some demographic questions: age, gender, experience in programming, main programming language used, original programming education, and main programming activity today.
\end{enumerate}

The survey was mostly based on multiple choice questions, on a scale of 1--5, where 1 means ``not at all important'' or ``strongly disagree'' and 5 means ``very important'' or ``strongly agree''.
In addition we used open questions where participants could expand on issues that they found missing in the multiple choice questions.

\subsection{Survey Execution}

The survey was implemented on the Google Forms platform.
Requests to participate in the survey were distributed both locally and globally.
Locally the survey was posted to student WhatsApp groups and Facebook groups, and sent directly to colleagues.
Globally the invitation to participate was posted to a few Reddit groups.
Consent to participate was explicitly implied by continuing from the introductory page to the survey questions.
No identifying information was collected, and participants were not compensated for taking the survey.

A total of 367 people responded to the survey.
89\% were men, 5.4\% were women, and a similar fraction preferred not to identify or gave a different answer.
While this is highly skewed, it appears to be representative of the developer community as reflected by the Stack Overflow developer survey%
\footnote{https://survey.stackoverflow.co/2022\#developer-profile-demographics}.
Of the 265 who answered the question about where they are from, 85 were from the United States,
41 from Israel,
28 from the United Kingdom,
19 from Germany,
and 8 each from the Netherlands and Australia.
The rest came from 34 other countries.
The ages of three quarters of the respondents were between 20 and 40, with 22\% being in the range 30--35.
Another quarter were above 40, with 8\% even above 50.

Concerning programming experience, 51\% had learned to program in a university or college, and 31\% studied independently.
10\% learned at work, in a bootcamp, or during military service.
88\% reported that their main programming activity is done at work.
8\% said it was done in independent projects, and the rest programmed in a research setting.
Their years of experience were rather skewed:
36\% had 0--5 years of experience, and another 24\% had 6-10 years.
13\% claimed 21 years of experience or more.
The combination of programming at work and the years of experience indicate that the vast majority of participants are professional developers.
The main programming languages used by participants were Python, TypeScript, and JavaScript, with 24--27\% naming them.
C\# was named by 18\%, Java by 13\%, and ten additional languages by 3--12\%.

\subsection{Survey Results}

\subsubsection{Use and Utility of Function Header Comments}~\\[-2mm]

We started the survey with a general question about actual practice:
does the participant's current project have documented functions?
A bit more than a quarter of the participants answered that no, functions are not documented.
A a bit less than half said that most functions do not have header comments but some do, and a bit more than a quarter said that most do.

\vspace{1mm}
\noindent\begin{tabular}{@{}p{0.98\columnwidth}}
\hline

\rule{0pt}{2ex}Does the codebase/project you currently work on have documented functions?\\

\begin{tikzpicture}
\pie[radius=1.3, color={red!70,yellow!50,blue!40}, sum=auto,
    text=pin, after number=\%, font=\small\sffamily]
    {26/\parbox{25mm}{\textsf no documented\\ functions},
     47/\parbox{20mm}{\textsf most\\ functions\\undocumented},
     27/\parbox{30mm}{\textsf most functions\\ have documentation} }
\end{tikzpicture}

\\[-4mm] \hline
\end{tabular}
\vspace{1mm}

The next question was about their personal behavior, and specifically where do they start from when they need to use a function they did not write.
More than half answered that they first read the header documentation (including its signature).
More than a third said they first read the code.
Few said they turn to external documentation or mentioned other options, such as tests.

\vspace{1mm}
\noindent\begin{tabular}{@{}p{0.98\columnwidth}}
\hline

\rule{0pt}{2ex}When you need to use a function you didn't write for the first time, what do you read first in order to understand the function usage?\\

\begin{tikzpicture}
\pie[radius=1.3, color={blue!50,green!30,red!70,yellow!50,black!40}, sum=auto,
    text=pin, after number=\%, font=\small\sffamily]
    {6/\textsf{external documentation},
     56/\parbox{20mm}{\textsf{header\\documentation}},
     36/\textsf{the code},
     1/\textsf{tests},
     1/\textsf{other} }
\end{tikzpicture}

\\[-4mm] \hline
\end{tabular}
\vspace{1mm}

\hide{
Responses under "Other":
    • Usually I let CoPilot infer it, otherwise StackOverflow or the wiki if I want usage examples and not just the description of parameters.
    • Both function code and usage in code and tests.
    • Follow through the chain of methods as sometimes large functions are divided into smaller code blocks for various purposes.
    • Other uses of the function in the code.
    • Examples that already use it

Responses that mentioned the function's name or arguments were merged into the option "The function's header documentation (including its signature)", those that mentioned tests into "Tests" and those that mentioned Googling/external documentation into "External documentation for the code".
}

Based on these answers, we asked those who said they start with the header comments whether this was enough.
The majority agreed that this was indeed the case, and that it is not necessary to also read the code.

\vspace{1mm}
\noindent\begin{tabular}{@{}p{44mm}p{36mm}}
\hline

\rule{0pt}{2ex}\raggedright Usually, one does not need to read the actual code to use the function, and can read only its header documentation to use it
[scale: 1=strongly disagree to 5=strongly agree] &

\adjustbox{valign=t}{\begin{tikzpicture}[scale=0.5]
 \begin{axis}[
   font=\sffamily,
   label style={font=\LARGE\sffamily}, tick label style={font=\LARGE\sffamily},
   width=170pt,
   height=100pt,
   scale only axis=true,
   ybar,
   bar width=15pt,
   xtick=data,
   enlarge x limits=0.15,
   ylabel = {Percents},
   ymin=0,ymax=40,
   ytick={0,10,20,30,40,50},
  ]
  \addplot[draw=blue!60,fill=blue!60,mark=none] coordinates {(1,8.3) (2,12) (3,22.1) (4,37.3) (5,20.3) };
 \end{axis}
\end{tikzpicture}}

\\ \hline
\end{tabular}
\vspace{2mm}

Alternatively, those who said they do not start with the header documentation were asked why.
More than half chose the option that code gives better understanding.
Around 40\% each selected criticisms of header documentation (not good or not up to date --- a concern that has also been noted in other research \cite{head18,levy21}) or that reading the code was simpler.
Few entered their own reasons --- that header documentation was missing, or that external documentation was better.

\vspace{1mm}
\noindent\begin{tabular}{@{}p{0.98\columnwidth}}
\hline

\rule{0pt}{2ex}Why do you start from the code/external documentation and not the function header documentation?\\

\multicolumn{1}{@{}r@{}}{
\adjustbox{valign=t}{\begin{tikzpicture}[scale=0.5]
 \begin{axis}[
   font=\sffamily,
   label style={font=\LARGE\sffamily}, tick label style={font=\sffamily\LARGE},
   width=150pt,
   height=100pt,
   scale only axis=true,
   xbar,
   bar width=10pt,
   ytick=data,
   yticklabel style={font=\sffamily\LARGE},
  yticklabels={Header docs often not up to date,Reading code is simpler,Header docs are not good,Code gives better understanding},
   enlarge y limits=0.2,
   xlabel = {Percents},
   xmin=0,xmax=60,
   xtick={0,10,20,30,40,50,60},
   legend pos=south east
  ]
   \addplot[fill=blue,color=blue!60,mark=none] coordinates {(39.6,1) (43,2) (44.3,3) (55,4) };
 \end{axis}
\end{tikzpicture}}}

\\ \hline
\end{tabular}
\hide{
24 (16.1\%) user-inputted answers: 
    • 6 (4\%) of those mention the reason that there is almost no function documentation or none at all.
    • 4 (2.6\%) of those the reason is "External documentation is usually more elaborate/has examples".
}

In addition, they were asked whether they later do read the header documentation.
The answers indicate that about a third sometimes do (answered 3), but half tend not to (1 or 2).
Only a small minority tend to read header documentation later (4 or 5).

\vspace{1mm}
\noindent\begin{tabular}{@{}p{44mm}p{36mm}}
\hline

\rule{0pt}{2ex}\raggedright How often do you go back to reading the function header documentation in later stages of understanding its usage?
[scale: 1=never to 5=always] &

\adjustbox{valign=t}{\begin{tikzpicture}[scale=0.5]
 \begin{axis}[
   font=\sffamily,
   label style={font=\LARGE\sffamily}, tick label style={font=\LARGE\sffamily},
   width=170pt,
   height=100pt,
   scale only axis=true,
   ybar,
   bar width=15pt,
   xtick=data,
   enlarge x limits=0.15,
   ylabel = {Percents},
   ymin=0,ymax=40,
   ytick={0,10,20,30,40,50},
  ]
  \addplot[draw=blue!60,fill=blue!60,mark=none] coordinates {(1,20.1) (2,29.5) (3,34.9) (4,11.4) (5,4) };
 \end{axis}
\end{tikzpicture}}
\\

\hline
\end{tabular}
\vspace{2mm}

Finally, all the participants were asked about their perception of the usefulness of function header documentation.
The majority thought they were useful.

\vspace{1mm}
\noindent\begin{tabular}{@{}p{44mm}p{36mm}}
\hline

\rule{0pt}{2ex}\raggedright How useful do you think functions' header documentation is?
[scale: 1=not at all to 5=very much] &

\adjustbox{valign=t}{\begin{tikzpicture}[scale=0.5]
 \begin{axis}[
   font=\sffamily,
   label style={font=\LARGE\sffamily}, tick label style={font=\LARGE\sffamily},
   width=170pt,
   height=100pt,
   scale only axis=true,
   ybar,
   bar width=15pt,
   xtick=data,
   enlarge x limits=0.15,
   ylabel = {Percents},
   ymin=0,ymax=40,
   ytick={0,10,20,30,40,50},
  ]
  \addplot[draw=blue!60,fill=blue!60,mark=none] coordinates {(1,4.7) (2,13.7) (3,21.9) (4,28.5) (5,31.2) };
 \end{axis}
\end{tikzpicture}}

\\ \hline
\end{tabular}
\vspace{2mm}

\subsubsection{The Elements of Function Header Comments}~\\[-2mm]

Header comments usually include several parts.
We specifically considered four: descriptions of the parameters, of the return value, of the internal structure, and usage examples.
The justifications for choosing them are as follows: Parameters and return value are universally considered the basic requirement to define an API; Inner functions represent implementation details; Examples are known to be desirable from past work.

All the participants were asked: In your opinion, how useful are the following parts of function header documentation?
The answers to these questions are shown in orange in the following graphs, on a scale of 1=not useful to 5=very useful.

Participants who had previously answered that they read header comments first were also asked: When reading function header documentation, how much do you focus on [these parts]?
The answers are shown in blue, on a scale of 1=not at all to 5=very much.

The results indicate that a strong correlation exists between the perception of usefulness by all participants and the focus by those participants who indeed read header documentation first.
The output definition and return value are valued most, closely followed by usage examples.
Explanations of the function parameters are also thought useful, but to a somewhat lesser degree.
The function's internal methods, on the other hand, are not considered useful or focused upon.
This is as can be expected when the header comments are used to learn how to use the function, as opposed to maintaining it.

\vspace{1mm}
\noindent\begin{tabular}{@{}p{44mm}p{36mm}}
\hline

\rule{0pt}{2ex}\raggedright Orange: How useful are, and blue: How much do you focus on: Explanations of the function's parameters &

\adjustbox{valign=t}{\begin{tikzpicture}[scale=0.5]
 \begin{axis}[
   font=\sffamily,
   label style={font=\LARGE\sffamily}, tick label style={font=\LARGE\sffamily},
   width=170pt,
   height=100pt,
   scale only axis=true,
   ybar,
   bar width=10pt,
   xtick=data,
   enlarge x limits=0.15,
   ylabel = {Percents},
   ymin=0,ymax=40,
   ytick={0,10,20,30,40,50},
  ]
  \addplot[draw=orange!50,fill=orange!50,mark=none] coordinates {(1,4.4) (2,11.2) (3,18.6) (4,27.7) (5,38.1) };
  \addplot[draw=blue!60,fill=blue!60,mark=none] coordinates {(1,6.9) (2,15.2) (3,21.7) (4,28.1) (5,28.1) };
 \end{axis}
\end{tikzpicture}}

\\ \hline

\rule{0pt}{2ex}\raggedright Orange: How useful are, and blue: How much do you focus on: Output definition and return value &

\adjustbox{valign=t}{\begin{tikzpicture}[scale=0.5]
 \begin{axis}[
   font=\sffamily,
   label style={font=\LARGE\sffamily}, tick label style={font=\LARGE\sffamily},
   width=170pt,
   height=125pt,
   scale only axis=true,
   ybar,
   bar width=10pt,
   xtick=data,
   enlarge x limits=0.15,
   ylabel = {Percents},
   ymin=0,ymax=50,
   ytick={0,10,20,30,40,50},
  ]
  \addplot[draw=orange!50,fill=orange!50,mark=none] coordinates {(1,3.6) (2,9.1) (3,14.3) (4,28) (5,45.1) };
  \addplot[draw=blue!60,fill=blue!60,mark=none] coordinates {(1,3.7) (2,8.8) (3,13.4) (4,28.1) (5,46.1) };
 \end{axis}
\end{tikzpicture}}

\\ \hline

\rule{0pt}{2ex}\raggedright Orange: How useful are, and blue: How much do you focus on: Explanation of the function's inner methods &

\adjustbox{valign=t}{\begin{tikzpicture}[scale=0.5]
 \begin{axis}[
   font=\sffamily,
   label style={font=\LARGE\sffamily}, tick label style={font=\LARGE\sffamily},
   width=170pt,
   height=100pt,
   scale only axis=true,
   ybar,
   bar width=10pt,
   xtick=data,
   enlarge x limits=0.15,
   ylabel = {Percents},
   ymin=0,ymax=40,
   ytick={0,10,20,30,40,50},
  ]
  \addplot[draw=orange!50,fill=orange!50,mark=none] coordinates {(1,28.8) (2,34) (3,21.9) (4,8.8) (5,6.6) };
  \addplot[draw=blue!60,fill=blue!60,mark=none] coordinates {(1,37) (2,35.2) (3,18.1) (4,7.9) (5,1.9) };
 \end{axis}
\end{tikzpicture}}

\\ \hline

\rule{0pt}{2ex}\raggedright Orange: How useful are, and blue: How much do you focus on: Examples of usage &

\adjustbox{valign=t}{\begin{tikzpicture}[scale=0.5]
 \begin{axis}[
   font=\sffamily,
   label style={font=\LARGE\sffamily}, tick label style={font=\LARGE\sffamily},
   width=170pt,
   height=125pt,
   scale only axis=true,
   ybar,
   bar width=10pt,
   xtick=data,
   enlarge x limits=0.15,
   ylabel = {Percents},
   ymin=0,ymax=50,
   ytick={0,10,20,30,40,50},
  ]
  \addplot[draw=orange!50,fill=orange!50,mark=none] coordinates {(1,5.2) (2,7.4) (3,14.6) (4,27.5) (5,45.3) };
  \addplot[draw=blue!60,fill=blue!60,mark=none] coordinates {(1,5.5) (2,8.8) (3,24) (4,25.8) (5,35.9) };
 \end{axis}
\end{tikzpicture}}

\\ \hline

\end{tabular}
\vspace{2mm}

Participants who said they read header documentation were then asked what other parts of the documentation they look for beyond the four we asked about explicitly.
A bit more than a third answered this question.
About 20\% each (of those who answered) mentioned a general description of the function and types.
Several respondents also mentioned other data items, such as exceptions, side effects, and more.
\hide{
Are there other parts you look for in the documentation?
73 Responses. Meaningful responses were categorized according to what was mentioned. Category count for answers associated with them were as follows:
Category                        Count
Function description/behavior   14
Types                           13
Exceptions/throws               8
Side effects                    7
Other/external documentation    5
Complexity                      4
Tests                           3
Reason code exists              2
}

Finally, all participants were asked a follow-up open question concerning information that is often lacking from function header documentation.
The most commonly cited missing element by far was examples of usage, mentioned by 42\%.
The next three issues, mentioned by 16\% each, were exceptions, side-effects, and being up to date.

\vspace{1mm}
\noindent\begin{tabular}{@{}p{0.98\columnwidth}}
\hline

\rule{0pt}{2ex}What information is usually lacking in functions' header documentation? (open question)\\

\multicolumn{1}{@{}r@{}}{
\adjustbox{valign=t}{\begin{tikzpicture}[scale=0.5]
 \begin{axis}[
   font=\sffamily,
   label style={font=\LARGE\sffamily}, tick label style={font=\LARGE\sffamily},
   width=150pt,
   height=200pt,
   scale only axis=true,
   xbar,
   bar width=10pt,
   ytick=data,
   yticklabel style={font=\sffamily\LARGE},
  yticklabels={Performance,Parameter descriptions,Description of return values,Types,Being up to date,Side effects,Exceptions,Example/usage},
   enlarge y limits=0.1,
   xlabel = {Percents},
   xmin=0,xmax=50,
   xtick={0,10,20,30,40,50,60},
   legend pos=south east
  ]
   \addplot[fill=blue,color=blue!60,mark=none] coordinates {(5,1) (5,2) (7,3) (9,4) (16,5) (16,6) (16,7) (42,8) };
 \end{axis}
\end{tikzpicture}}
}\\
\hline

\end{tabular}
\vspace{2mm}

\pagebreak

\subsubsection{Writing Function Header Comments}~\\[-1mm]

The questions about writing header comments were preceded by a note that participants who do not do so may not answer this part.
Nevertheless, 350 participants (95\%) did answer these questions.

The first two questions in this part of the survey exhibit somewhat inconsistent behavior of the participants:
most of them think that writing function header documentation is worth the time it requires, but many often do not actually write such documentation.
Note that the question specifically noted that we are referring to functions that might be used by other developers in the future, not internal private functions.

\vspace{1mm}
\noindent\begin{tabular}{@{}p{44mm}p{36mm}}
\hline

\rule{0pt}{2ex}\raggedright Do you think writing function header documentation is worth the time it requires?
[scale: 1=not at all to 5=very much] &

\adjustbox{valign=t}{\begin{tikzpicture}[scale=0.5]
 \begin{axis}[
   font=\sffamily,
   label style={font=\LARGE\sffamily}, tick label style={font=\LARGE\sffamily},
   width=170pt,
   height=100pt,
   scale only axis=true,
   ybar,
   bar width=15pt,
   xtick=data,
   enlarge x limits=0.15,
   ylabel = {Percents},
   ymin=0,ymax=40,
   ytick={0,10,20,30,40,50},
  ]
  \addplot[draw=blue!60,fill=blue!60,mark=none] coordinates {(1,8.9) (2,13.7) (3,20.6) (4,25.4) (5,31.4) };
 \end{axis}
\end{tikzpicture}}

\\ \hline

\rule{0pt}{2ex}\raggedright How often do you write documentation to your functions that other people might use in the future?
[scale: 1=never to 5=always] &

\adjustbox{valign=t}{\begin{tikzpicture}[scale=0.5]
 \begin{axis}[
   font=\sffamily,
   label style={font=\LARGE\sffamily}, tick label style={font=\LARGE\sffamily},
   width=170pt,
   height=75pt,
   scale only axis=true,
   ybar,
   bar width=15pt,
   xtick=data,
   enlarge x limits=0.15,
   ylabel = {Percents},
   ymin=0,ymax=30,
   ytick={0,10,20,30,40,50},
  ]
  \addplot[draw=blue!60,fill=blue!60,mark=none] coordinates {(1,8.3) (2,25.1) (3,21.1) (4,24.3) (5,21.1) };
 \end{axis}
\end{tikzpicture}}

\\ \hline
\end{tabular}
\vspace{1mm}

Expecting that this may be the case, we asked those who tend not to write function header documentation (identified as those who answered 1 or 2) why they behave this way.
The most common answer given, by more than a third of the participants, is that in fact they feel such documentation is unimportant (we include here respondents who mentioned alternatives like good names, especially for short simple functions, or external documentation).
Around a sixth each answered that it requires too much time, that self-explanatory code is more important, and that it is not commonly done in their team.
Of these, the answer that self-explanatory code is more important was not suggested by us, but written in under ``other''.
Additional answers given under ``other'' include
the code being for internal use only,
the danger of the documentation not being updated,
and that it interrupts the workflow and is then forgotten.

\vspace{1mm}
\noindent\begin{tabular}{@{}p{0.98\columnwidth}}
\hline

\rule{0pt}{2ex}Why do you avoid writing function header documentation? \\

\multicolumn{1}{@{}r@{}}{
\adjustbox{valign=t}{\begin{tikzpicture}[scale=0.5]
 \begin{axis}[
   font=\sffamily,
   label style={font=\LARGE\sffamily}, tick label style={font=\LARGE\sffamily},
   width=150pt,
   height=200pt,
   scale only axis=true,
   xbar,
   bar width=10pt,
   ytick=data,
   yticklabel style={font=\sffamily\LARGE},
  yticklabels={Other,Tests explain the code better [other],The functions are for internal use [other],Danger that it won't be updated [other],It's important but uncommon in my team,Self explanatory code more important [other],It's important but requires too much time,I feel it's unimportant},
   enlarge y limits=0.1,
   xlabel = {Percents},
   xmin=0,xmax=40,
   xtick={0,10,20,30,40,50,60},
   legend pos=south east
  ]
   \addplot[fill=blue,color=blue!60,mark=none] coordinates {(6.9,1) (2.5,2) (3.1,3) (3.7,4) (15,5) (16.9,6) (17.5,7) (34.4,8) };
 \end{axis}
\end{tikzpicture}}}

\\ \hline
\end{tabular}
\vspace{2mm}
\hide{
additional categories in ``other'':
for internal use only
danger of not being updated
two said not in workflow (example: It requires a separate "voice" / "flow", so I don't do it initially, and sometimes don't come back to it.)
unimportant included mention of alternatives: good names especially for short simple functions or external documentation
to much time may be qualified for short simple functions (example: When the header would be longer than the code it documents, it's not worth the time -- someone else would rather just read the code instead.)
}

Those who answered 3--5 to the question of how often they document functions were asked how much they elaborate on different elements in this documentation (using a scale of 1=not at all to 5=a lot).
The answers indicated that there was some elaboration on parameters, and less on examples --- in both cases much less than reflecting the perceived importance of these elements as reported above.
There was a strong tendency not to elaborate on internal functions, which we saw are indeed not perceived as important.

\vspace{1mm}
\noindent\begin{tabular}{@{}p{44mm}p{36mm}}
\hline

\rule{0pt}{2ex}\raggedright How much do you elaborate in explaining the function's parameters in its header documentation? &

\adjustbox{valign=t}{\begin{tikzpicture}[scale=0.5]
 \begin{axis}[
   font=\sffamily,
   label style={font=\LARGE\sffamily}, tick label style={font=\LARGE\sffamily},
   width=170pt,
   height=100pt,
   scale only axis=true,
   ybar,
   bar width=15pt,
   xtick=data,
   enlarge x limits=0.15,
   ylabel = {Percents},
   ymin=0,ymax=40,
   ytick={0,10,20,30,40,50},
  ]
  \addplot[draw=blue!60,fill=blue!60,mark=none] coordinates {(1,6.3) (2,19.8) (3,34.9) (4,29.8) (5,9.1) };
 \end{axis}
\end{tikzpicture}}

\\ \hline

\rule{0pt}{2ex}\raggedright How much do you elaborate in explaining the function's inner methods in its header documentation? &

\adjustbox{valign=t}{\begin{tikzpicture}[scale=0.5]
 \begin{axis}[
   font=\sffamily,
   label style={font=\LARGE\sffamily}, tick label style={font=\LARGE\sffamily},
   width=170pt,
   height=135pt,
   scale only axis=true,
   ybar,
   bar width=15pt,
   xtick=data,
   enlarge x limits=0.15,
   ylabel = {Percents},
   ymin=0,ymax=54,
   ytick={0,10,20,30,40,50},
  ]
  \addplot[draw=blue!60,fill=blue!60,mark=none] coordinates {(1,50.4) (2,24.6) (3,16.7) (4,6.3) (5,2) };
 \end{axis}
\end{tikzpicture}}

\\ \hline

\rule{0pt}{2ex}\raggedright How much do you elaborate in explaining the function's examples of usage in its header documentation? &

\adjustbox{valign=t}{\begin{tikzpicture}[scale=0.5]
 \begin{axis}[
   font=\sffamily,
   label style={font=\LARGE\sffamily}, tick label style={font=\LARGE\sffamily},
   width=170pt,
   height=75pt,
   scale only axis=true,
   ybar,
   bar width=15pt,
   xtick=data,
   enlarge x limits=0.15,
   ylabel = {Percents},
   ymin=0,ymax=30,
   ytick={0,10,20,30,40,50},
  ]
  \addplot[draw=blue!60,fill=blue!60,mark=none] coordinates {(1,15.7) (2,28.4) (3,28.8) (4,16.9) (5,19.2) };
 \end{axis}
\end{tikzpicture}}

\\ \hline
\end{tabular}
\vspace{1mm}

Finally, we asked all participants how often they update the documentation compared with updating the function's code.
The answers indicate that developers tend not to update the documentation as much as may be desired, justifying the reservations reported above concerning documentation that may not be up to date.
However, we note that Malik et al.\ found that the fraction of comments that were not updated in 3 of the projects they studied were lower than would be expected by our survey, ranging from 18\% to 33\%; in only one project it was a high 62\% \cite{malik08}.
Likewise, Chen et al., in a study on identifying comments scope, estimated that only around 13\% of the comments in their sample were outdated \cite{chen19}.
This discrepancy might be explained by code updates that do not require a change in the header comment, for example because they just fix a bug.

\vspace{1mm}
\noindent\begin{tabular}{@{}p{44mm}p{36mm}}
\hline

\rule{0pt}{2ex}\raggedright How often do you update the function header documentation, compared to the updates in the function's code?
[scale: 1=never to 5=in every code update] &

\adjustbox{valign=t}{\begin{tikzpicture}[scale=0.5]
 \begin{axis}[
   font=\sffamily,
   label style={font=\LARGE\sffamily}, tick label style={font=\LARGE\sffamily},
   width=170pt,
   height=100pt,
   scale only axis=true,
   ybar,
   bar width=15pt,
   xtick=data,
   enlarge x limits=0.15,
   ylabel = {Percents},
   ymin=0,ymax=40,
   ytick={0,10,20,30,40,50},
  ]
  \addplot[draw=blue!60,fill=blue!60,mark=none] coordinates {(1,16.9) (2,33.4) (3,22.8) (4,12.1) (5,14.8) };
 \end{axis}
\end{tikzpicture}}

\\ \hline
\end{tabular}
\vspace{1mm}

\subsubsection{Using Tools and Documentation Templates}~\\[-1mm]

A couple of questions concerned the use of auto-generated documentation and its quality.
The first showed that half the participants never use such tools.

\vspace{1mm}
\noindent\begin{tabular}{@{}p{44mm}p{36mm}}
\hline

\rule{0pt}{2ex}\raggedright How often do you use tools for automatic documentation generation (e.g.\ docstring generation tools built into IDEs)?
[scale: 1=never to 5=very often] &

\adjustbox{valign=t}{\begin{tikzpicture}[scale=0.5]
 \begin{axis}[
   font=\sffamily,
   label style={font=\LARGE\sffamily}, tick label style={font=\LARGE\sffamily},
   width=170pt,
   height=135pt,
   scale only axis=true,
   ybar,
   bar width=15pt,
   xtick=data,
   enlarge x limits=0.15,
   ylabel = {Percents},
   ymin=0,ymax=54,
   ytick={0,10,20,30,40,50},
  ]
  \addplot[draw=blue!60,fill=blue!60,mark=none] coordinates {(1,51.2) (2,15.5) (3,9.5) (4,10.9) (5,12.8) };
 \end{axis}
\end{tikzpicture}}

\\ \hline
\end{tabular}
\vspace{1mm}

The second question actually had two versions.
All participants were asked:
How much do you think automatically generated documentation should be changed?
This is shown in orange in the graph, on a scale of 1=not at all (useful as it is) to 5=a lot (useful only if edited). 
The second version was directed only to participants who do use automatic generation, who were asked:
How much do you edit or add to the generated docstring?
This is shown in blue, with a scale of 1=I don't edit anything to 5=I edit a lot. 
The result show a wide spread of opinions, with a moderate correlation between thinking generated documentation should be edited and actually doing it.

\vspace{1mm}
\noindent\begin{tabular}{@{}p{44mm}p{36mm}}
\hline

\rule{0pt}{2ex}\raggedright Orange: How much should auto-generated docstrings be changed?\\ Blue: How much do you edit or add to them? &

\adjustbox{valign=t}{\begin{tikzpicture}[scale=0.5]
 \begin{axis}[
   font=\sffamily,
   label style={font=\LARGE\sffamily}, tick label style={font=\LARGE\sffamily},
   width=170pt,
   height=100pt,
   scale only axis=true,
   ybar,
   bar width=10pt,
   xtick=data,
   enlarge x limits=0.15,
   ylabel = {Percents},
   ymin=0,ymax=40,
   ytick={0,10,20,30,40,50},
  ]
  \addplot[draw=orange!50,fill=orange!50,mark=none] coordinates {(1,11.6) (2,12.3) (3,24.9) (4,18.2) (5,33) };
  \addplot[draw=blue!60,fill=blue!60,mark=none] coordinates {(1,23) (2,14.4) (3,14.9) (4,21.3) (5,26.4) };
 \end{axis}
\end{tikzpicture}}

\\ \hline
\end{tabular}
\vspace{2mm}

\subsubsection{Function Header Comments Guidelines}~\\[-1mm]

Finally, we asked participants whether they are required to document the functions they write.
Nearly two thirds of the respondents said no, and a quarter said yes but that there were no explicit guidelines.
Less than 8\% said they had explicit guidelines to document functions.
A few participants added a comment to the effect that they are required to document only complex functions, or public/main functions.
Two said that documentation is forbidden.

\vspace{1mm}
\noindent\begin{tabular}{@{}p{0.98\columnwidth}}
\hline

\rule{0pt}{2ex}Are you required by your team's/company's guidelines to document the functions you write?\\

\hspace*{22mm}\begin{tikzpicture}
\pie[radius=1.3, color={red,orange!70,blue!30,green!70,yellow!50,black!40}, sum=auto,
    text=pin, after number=\%, font=\small\sffamily]
    {0.5/\textsf{forbidden},
     62.3/\textsf{no},
     25.1/\textsf{yes but not explicit},
     7.7/\textsf{explicit guidelines},
     2.5/\textsf{if complex/public},
     1.9/\textsf{other} }
\end{tikzpicture}

\\[-4mm] \hline
\end{tabular}
\vspace{1mm}
\hide{
some comments given here:
5 said that if complex
4 said that if public/main
two mentioned good function names as the alternative
one mentioned tests: TDD gives us executable docs that are always current
two explicitly stated that function header documentation is not allowed and considered a bad practice
two said they wrote the guidelines...
}

A follow-up question for those who said they have documentation guidelines asked what these guidelines included.
The most common elements were a summary or general explanation of the function (86\%), and an explanation of its parameters (56\%).
Only 20\% were required to provide examples, which were identified above as a part that is often missing.

\vspace{1mm}
\noindent\begin{tabular}{@{}p{0.98\columnwidth}}
\hline

\rule{0pt}{2ex}If your group has guidelines, what do they require? \\

\multicolumn{1}{@{}r@{}}{
\adjustbox{valign=t}{\begin{tikzpicture}[scale=0.5]
 \begin{axis}[
   font=\sffamily,
   label style={font=\LARGE\sffamily}, tick label style={font=\LARGE\sffamily},
   width=180pt,
   height=120pt,
   scale only axis=true,
   xbar,
   bar width=10pt,
   ytick=data,
   yticklabel style={font=\sffamily\LARGE},
  yticklabels={Explanation of function's inner methods,Examples of usage,Using an automatic template/docstring,Explanation of parameters,Short summary/general explanation},
   enlarge y limits=0.15,
   xlabel = {Percents},
   xmin=0,xmax=100,
   xtick={0,20,40,60,80,100},
   legend pos=south east
  ]
   \addplot[fill=blue,color=blue!60,mark=none] coordinates {(8.9,1) (18.7,2) (19.5,3) (56.1,4) (86.2,5) };
 \end{axis}
\end{tikzpicture}}}

\\ \hline
\end{tabular}
\vspace{2mm}

Alternatively, those who did not have documentation guidelines were asked why not (this was an open question).
The most common answer, given by 31\%, was that code should be self-documenting.
24\% cited a startup culture, including not having enough time or having a small team.
Many others had opposing opinions on the matter:
19\% noted that they do not want documentation, while 14\% were critical of the fact that documentation guidelines do not exist.
Some of these were downright disparaging of the company, using ``they'' to describe managers who make such decisions without understanding their consequences.

\vspace{1mm}
\noindent\begin{tabular}{@{}p{0.98\columnwidth}}
\hline

\rule{0pt}{2ex}What is the reason your team/company does not require writing function documentation? \\

\multicolumn{1}{@{}r@{}}{
\adjustbox{valign=t}{\begin{tikzpicture}[scale=0.5]
 \begin{axis}[
   font=\sffamily,
   label style={font=\LARGE\sffamily}, tick label style={font=\LARGE\sffamily},
   width=150pt,
   height=120pt,
   scale only axis=true,
   xbar,
   bar width=10pt,
   ytick=data,
   yticklabel style={font=\sffamily\LARGE},
  yticklabels={Documentation is hard to maintain,Criticism on not requiring documentation,Don't want documentation,Startup culture,Code should be self-documenting},
   enlarge y limits=0.15,
   xlabel = {Percents},
   xmin=0,xmax=33,
   xtick={0,10,20,30,40,50},
   legend pos=south east
  ]
   \addplot[fill=blue,color=blue!60,mark=none] coordinates {(5.8,1) (14,2) (19.1,3) (23.8,4) (30.8,5) };
 \end{axis}
\end{tikzpicture}}}

\\ \hline
\end{tabular}
\vspace{1mm}
\hide{
criticism, some downright disparaging of the company. using ``they'' about managers who decide but don't get it.
startup culture includes not enough time and small teams
names+simple=not needed includes exercise individual judgement about when needed; many quote code should document itself, some quote clean code explicitly, some mention splitting functions that need doc, some disagree
no ranges from not necessary to not discussed; includes doc system or modules, not individual functions
}

For example, one of the detailed comments we received was:
``1. Concise and explanatory code is more readable than documentation; 2. Documentation needs to be maintained. When it's not maintained it can be harmful: it gets out of sync and the docs turn to be unreliable. When it's unreliable people will ignore it altogether, and the handful of functions that really need to be elaborated can't be documented anymore.''
Another wrote:
``Comments are a last resort if you can't make the code simple enough to understand on its own. By making comments rarer you know that if there's a comment it's probably worth reading.''
\hide{
"Lots of experiments leading to a lot of throwaway code. ROI for documenting that code would be low. Documentation is usually added when code becomes part of critical paths."
"We do require writing documentation, but at the system level and not the individual function level. We take much greater care in documenting configurations and the requirements for the system, then pass those off to QA for testing. We also have a client documentation team that we work with closely to give clients the tools they need to run our applications."
"Function documentation is mostly written as a time saver for copilot to generate the code in simple cases"
}


\subsection{Discussion: The Paradox}

The above results include many interesting observations and opinions about function header documentation.
But the most interesting and important in our mind is the following paradox.
On the one hand,
\begin{itemize}
    \item When using a function for the first time, 56\% turn to header documentation
    \item There is general agreement (answers of 4 or 5 on a scale of 1 to 5) that:
    \begin{itemize}
        \item Header documentation is useful (59.7\%)
        \item One can read only the header documentation and avoid reading the code (57.6\%)
        \item Writing function header documentation is worth the time it requires (56.8\%)
    \end{itemize}
\end{itemize}
But on the other hand
\begin{itemize}
    \item In only 27\% of the projects most functions have header documentation, and in 26\% none have
    \item Only 7.7\% of projects have explicit guidelines requiring header documentation, while 62.3\% have no guidelines of any kind
    \item Only 45.4\% of developers often write header documentation (answers 4--5), while 33.4\% often do not (answers 1--2) --- \emph{in functions that other people might use}
\end{itemize}
More specifically, the most significant disparity concerns examples.
72.8\% thought that usage examples are useful,
and 61.7\% said they focus on usage examples when reading header documentation.
At the same time 42\% said that usage examples are usually lacking from function header documentation,
and only 18.7\% had guidelines that require usage examples.
Likewise, 36.1\% elaborate on explaining usage examples when writing header documentation (answers 4--5), but 44.1\% do not (answers 1--2).
These results mirror those of Robillard from a survey of Microsoft employees, which also found examples to be the number one requirement from helpful API documentation \cite{robillard09}.

Luckily, this also points to what can be done to improve the situation.
The two recommendations that stand out are:
\begin{enumerate}
    \item Draft guidelines that require function header documentation to be written, especially for public and complex functions.
    \item Require that function header comments include examples of usage.
    This can be made easier using an example synthesizing approach such as that of Buse and Weimer \cite{buse12}.
    according to Meng et al., this should include both examples of simple use cases and recipes for more special use cases \cite{meng19}.
\end{enumerate}

\section{Mining GitHub for Data on Useless Header Comments}

One of the responses to the survey question of why do you not write header comments was ``Most documentation seems redundant given the naming and typing of arguments/return values''.
This identifies a major question hounding our analysis of the perception of header comments: are they in fact needed at all?
maybe all the pertinent information is already available in the function signature?

This also resonates with the practice of using documentation templates that include descriptions of parameters with no added value.
We've all seen examples such as this one (used as an example of what to avoid in an Oracle tutorial on using JavaDoc):
{\small\begin{verbatim}
  /**
  * Sets the tool tip text.
  *
  * @param text  the text of the tool tip
  */
  public void setToolTipText(String text) {
\end{verbatim}}
Obviously, the comment just repeats words that make up the signature.
We use this as inspiration for how to identify meaningless comments.
Given an implementation of this definition, we apply it to the files of multiple popular Python GitHub projects, and extract the distribution of ``meaninglessness''.
This shows the prevalence of writing meaningless documentation, and conversely --- the fraction of header comments that in fact contain additional information.

\subsection{Defining ``Meaningless'' Header Comments}

Header documentation is not the same as summarization.
A key element in documenting APIs is to document the interface itself (parameters and return values).
This is also what is required for contracts, and is implemented in various tools.
Our metric is aimed at this part of the documentation.

The metric for meaningless documentation is based on the idea that words that appear in the function signature do not add information when they are repeated in the header documentation.
This is essentially similar to the ``coherence coefficient'' suggested by Steidl et al.\ \cite{steidl13}.
However, our procedure to quantify it has several added features:
\begin{enumerate}

    \item For each function, we start with the full function signature (denoted $Sig$) and with its header documentation (denoted $Doc$).
    
    \item We process the documentation string as follows:
    \begin{enumerate}
        \item First we partition it into words.
        This is done in several steps.
        The first is to replace every sequence of non-alphanumeric characters with a space.
        Note that this already handles names that are written in snake\_case style.
        We also separate names written in camelCase into their constituent words based on their internal capitalization.
        The result of this step is a list of individual words in the documentation.
        \item From the list of words created in the previous step we extract the potentially meaningful words.
        These are words that satisfy two conditions:
        \begin{itemize}
            \item They are longer than a single character, and
            \item They are not a stop word.
            The list of stop words we used is `the', `an', `of', `at', `by', `in', `it', `on', `to', `that', `had', `for', `was', `were'.
            Note that certain words that are commonly included in lists of stop words are \emph{not} included here, as they may be very meaningful in the specification of functions.
            Examples are `and', `or', `is'.
        \end{itemize}
        \item The list of extracted words is denoted $W_{Doc}$, and the number of these words is $|W_{Doc}|$.
        Note that this is a bag and not a set: words may appear in it more than once.
    \end{enumerate}

    \item We process the function signature as follows:
    \begin{enumerate}
        \item First we partition it into words.
        This is done essentially in the same way as was done with the documentation string.
        \item We create a set of all the words that appear in the signature, denoted $W_{Sig}$.
        Note that this includes all of the following:
        \begin{itemize}
            \item Words that appear in the function name
            \item Words that appear in the function's parameters
            \item Words that appear in types of parameters and in the return type (if specified)
        \end{itemize}
        This is a set.
        There is no need to retain repetitions.
        
    \end{enumerate}

    \item For each word in the documentation, we check whether it appears in the signature, and if so declare it to be meaningless.
    This is done in two steps:
    \begin{enumerate}
        \item For each word $w \in W_{Doc}$ we check whether $w \in W_{Sig}$.
        This is straightforward inclusion of the word.
        \item If the word $w$ is not found, we check the opposite direction: for each word $v \in W_{Sig}$ we check whether $v \subset w$.
        This checks whether the signature includes shortened versions of words in the documentation.
        For example, `information' in the docstring would be counted as meaningless if `info' appears in the signature.
    \end{enumerate}
    The list of words from the documentation identified as meaningless is denoted $M$, and their number is $|M|$.
    This is a bag, with possible repetitions.

    \item Finally, we define the meaningless score of the function as $|M| / |W_{Doc}|$, namely the fraction of potentially meaningful words in the documentation that were found in the signature and are therefore actually meaningless.
    Thus, a large score (close to 1) indicates a rather meaningless documentation, as the information it provides was already given in the function's signature.
\end{enumerate}

Note that the above procedure defines ``meaningless'' based on the matching of individual words, and does not rely on semantics.
Thus it is actually a rather crude and preliminary definition.
We leave the task of designing more sophisticated metrics for the meaningful data provided by documentation to future work.

\subsection{The GitHub Dataset}

To check the actual prevalence of meaningless header documentation, 68 Python repositories were analyzed.
Repositories were chosen by popularity on GitHub, based on the Github Ranking as reflected in the Top 100 Stars in Python list%
\footnote{https://github.com/EvanLi/Github-Ranking/blob/master/Top100/Python.md}.
This is a dynamic list and gets updated frequently; the lists that were used are from December 2022 to January 2023.

Within each project we focused on the actual functional code and avoided test code, as identified by ``test'' (including plural or capitalized) or ``e2e'' appearing in the path.
A total of 21,650 files were analyzed.
510 of them (2.3\%) failed to be parsed (some because of encoding errors, some because of errors in the \code{docstring\_extractor} library used in the script).
The presented results are based on the remaining 21,140 files.

For each file we count the total number of functions in the file and the number of functions without any header documentation.
In addition we extract the docstrings from functions that do have one.
This is done using the function \code{get\_docstrings} from the \code{docstring\_extractor} Python library%
\footnote{https://pypi.org/project/docstring-extractor/}.
In addition, we use Python's \code{ast} (abstract syntax tree) library to extract function parameters and their types (if they exist).
We then apply the procedure described above to calculate the meaningless score of the docstring of each function, and collect these results.

\subsection{Results on Meaningless Header Comments}

We first calculate the average fraction of functions with no documentation.
To avoid giving more weight to large files, we first find the fraction of such functions in each file and then average over all the files from all the projects.
The resulting average fraction of functions with no documentation was 0.596.
This means that most functions are undocumented, as was also found in the survey.
The following results pertain to the other functions, which do have docstrings.

\begin{figure}\centering
    \includegraphics[width=0.97\columnwidth]{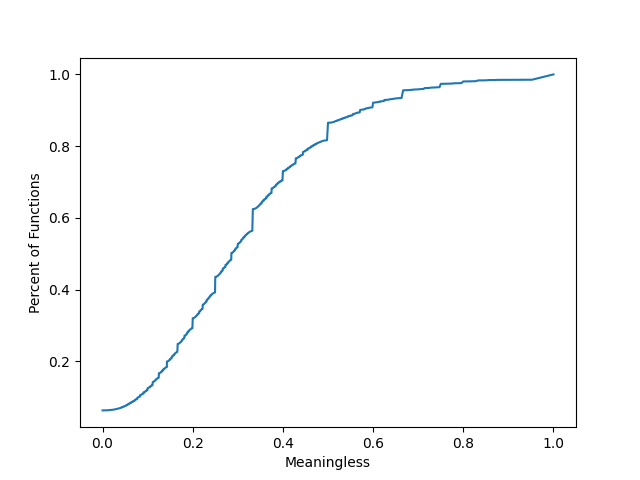}
    \caption{\label{fig:meaningless-all}
    Distribution of meaningless score for all functions that have docstrings.}
\end{figure}

Figure \ref{fig:meaningless-all} shows the CDF (cumulative distribution function) of the meaningless score of all the functions from all the projects.
For every value of meaningless score on the horizontal axis this shows the fraction of all functions whose header documentation has a meaningless score up to this value.
The graph indicates that the central 80\% of the distribution of functions have a meaningless score in the range 0.15--0.45.
In other words, for the vast majority of functions more than half the words in the header documentation are novel and do not appear in the function signature.
This is similar to the result obtained by Steidl et al., who found (in a sample of 5 projects) that 1--9\% of the header comments were ``trivial'', which they defined as having at least half of their words appear in the function signature.

\begin{figure}\centering
    \includegraphics[width=0.97\columnwidth]{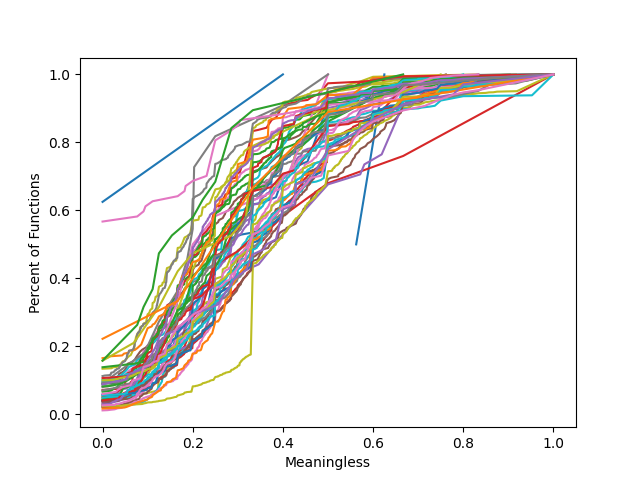}
    \caption{\label{fig:meaningless-by-proj}
    Distribution of meaningless score for the functions in different projects.}
\end{figure}

To verify that these results are indeed representative we plot the CDF of meaningless scores for each studied project separately.
All these graphs are shown together in Figure \ref{fig:meaningless-by-proj}.
As can be seen, the graphs for nearly all the projects roughly follow the same shape as the previous graph.

However, some of the projects exhibit different behaviors.
We manually checked two of them.
The one represented by the blue straight line at top left turned out to have code that was partially generated automatically from a .proto file (Google's protocol buffer file).
These automatically generated functions all had the default documentation string ``Missing associated documentation comment in .proto file''.
As none of these words appeared in the function signatures, they all got a meaningless score of 0, despite actually being meaningless.
The rest of the functions were constructors which received one argument, and had a header comment stating what this argument was.
These got a meaningless score of 0.4.

The project represented by the pink line below that had 34 functions called \code{main} with no arguments, which had verbose documentation including many examples.
They all got a meaningless score of 0, because the word ``main'' did not appear.

We conclude that the projects that exhibit strange behaviors indeed contain code that is not representative of normal programming by human developers.
However, there are only a small number of such projects, so their presence does not have a significant impact on the validity of the general results.

\section{Threats to Validity}

Documentation is a wide-ranging topic.
In the process of recruiting participants on reddit we received several comments to the effect that our survey did not and could not cover them all adequately.
For example, one participant noted that API documentation can never be complete because it is impossible to anticipate all the needs users might have.
Another noted that the documentation requirements in different languages could be different: in functional languages types are important, while in object oriented ones the parameter explanations are the most useful.
There were also reservations about the survey itself, e.g.\ that we did not ask explicitly about side effects being the most important thing missing in documentation.
While some participants added this and other considerations under ``other'', it stands to reason that more might have noted these considerations had they been presented explicitly.
The results are therefore probably biased against such considerations.
More generally, in various questions participants might have made different assumptions concerning what exactly we were asking about, leading to differences in answers.

Another concern is that our survey targeted developers and not projects.
While it would be surprising if we had multiple participants from the same project, this could in principle happen.
In any case, we cannot assume that the results are statistically representative of the whole industry, and results on percents of projects may be biased.
Likewise, the empirical data collected from GitHub was limited to projects in Python, and can not be assumed to even represent all projects in this language.

Regarding the empirical evidence we collected, since we used a very simple definition for ``meaningless'' score, it might be biased, and not capture the real essence of meaningless function header documentation.
We believe that this study showed the potential of this methodology, and thus additional effort needs to be put in the construction of a more sophisticated score for meaningless documentation.

As such, we see our results as indicative for the answers to our research questions, but concede that these results may be partial, as they might be biased by the projects we chose and people that answered our survey.
More participants in the survey, and more projects to be analyzed, might show a somewhat different picture, and are thus needed if one wishes to draw general conclusions from this study.

\section{Conclusions}

Returning to our research questions,
the answer to RQ1 is that header functions are not universally used.
About a quarter of our survey participants reported that most of the functions in their project are documented, and a similar fraction said that none are.
But more than half said they turn to header comments first when needing to understand a new function.
Concerning RQ2, the consensus was that output definition, examples, and parameters are all useful and important (in this order), while inner methods are not.
But in RQ3 we found that examples were most often missing from the documentation.
As for RQ4, based on our GitHub sample we found that while most functions do not have header comments, in those that do these comments do not just repeat information that is already available in the function signature.

To conclude, we believe that function header documentation has an important role in helping new developers to understand existing code, which is fulfilled only partially in practice.
This is due mainly to the large fraction of projects and functions which do not have such documentation, and to sub-par documentation which does not include examples.
Thus, programmers might tend to see header documentation as not worth the time investment in practice, both when reading it (in new functions they did not write) and when writing it (in projects they lead).
And indeed, we found a mismatch between developers' opinions that writing header comments is worth the effort and their actual behavior of not always doing so.

However, our results also indicate that this state of affairs can be improved with relatively little investment.
A key observation is that not all functions should be treated equally.
It is important to apply judgment, and only document functions that need it --- either because they are part of an API or because they are complex.
An easy test is to verify that you have what to say beyond the obvious from the function declaration.
In these cases, it is important to apply documentation guidelines which correspond to what developers want and need, especially regarding the provision of examples of usage.
In other cases, where header documentation is not really needed, it should be avoided, thereby reducing the overhead both for writers and for readers, and at the same time focusing attention on those functions that \emph{do} have documentation --- presumably for good reason.

\section*{Experimental Materials}

Experimental materials are available from Zenodo using DOI 10.5281/zenodo.10480939.

\bibliographystyle{myabbrv}
\bibliography{abbrv,se}

\begin{thebibliography}{10}\itemsep 0pt
\newcommand{\enquote}[1]{``#1''}
\providecommand{\url}[1]{\texttt{#1}}
\providecommand{\urlprefix}{URL }
\makeatletter
\let\doi\@undefined
\newcommand{\doi}[1]{\textsf{DOI: #1}}
\makeatother

\bibitem{aggarwal02}
K.~K. Aggarwal, Y.~Singh, and J.~Kumar, \enquote{\textsl{An integrated measure
  of software maintainability}}. In \textit{Proc.\ Ann.\ Reliability \&
  Maintainability Symp.}, pp.\ 235--241, Jan 2002,
  \doi{10.1109/RAMS.2002.981648}.

\bibitem{alsuhaibani21}
R.~S. Alsuhaibani, C.~D. Newman, M.~J. Decker, M.~L. Collard, and J.~I.
  Maletic, \enquote{\textsl{On the naming of methods: A survey of professional
  developers}}. In 43rd \textit{Intl.\ Conf.\ Softw.\ Eng.}, pp.\ 587--599, May
  2021, \doi{10.1109/ICSE43902.2021.00061}.

\bibitem{arab92}
M.~Arab, \enquote{\textsl{Enhancing program comprehension: Formatting and
  documenting}}. \textit{SIGPLAN Notices} \textbf{27(2)}, pp.\ 37--46, Feb
  1992, \doi{10.1145/130973.130975}.

\bibitem{berry16}
D.~M. Berry, K.~Czarnecki, M.~Antkiewicz, and M.~AbdElRazik,
  \enquote{\textsl{The problem of the lack of benefit of a document to its
  producer ({PotLoBoaDtiP})}}. In \textit{IEEE Intl.\ Conf.\ Softw.\ Sci.,
  Tech.\ \& Eng.}, pp.\ 37--42, Jun 2016, \doi{10.1109/SWSTE.2016.14}.

\bibitem{blinman05}
S.~Blinman and A.~Cockburn, \enquote{\textsl{Program comprehension:
  Investigating the effects of naming style and documentation}}. In 6th
  \textit{Australasian User Interface Conf.}, pp.\ 73--78, Jan 2005.

\bibitem{borstler16}
J.~B{\"o}rstler and B.~Paech, \enquote{\textsl{The role of method chains and
  comments in software readability and comprehension---an experiment}}.
  \textit{IEEE Trans.\ Softw.\ Eng.} \textbf{42(9)}, pp.\ 886--898, Sep 2016,
  \doi{10.1109/TSE.2016.2527791}.

\bibitem{buse12}
R.~P.~L. Buse and W.~Weimer, \enquote{\textsl{Synthesizing {API} usage
  exampes}}. In 34th \textit{Intl.\ Conf.\ Softw.\ Eng.}, pp.\ 782--792, Jun
  2012, \doi{10.1109/ICSE.2012.6227140}.

\bibitem{chen19}
H.~Chen, Y.~Huang, Z.~Liu, X.~Chen, F.~Zhou, and X.~Luo,
  \enquote{\textsl{Automatically detecting the scopes of source code
  comments}}. \textit{J.\ Syst.\ \& Softw.} \textbf{153}, pp.\ 45--63, Jul
  2019, \doi{10.1016/j.jss.2019.03.010}.

\bibitem{crosby02}
M.~E. Crosby, J.~Scholtz, and S.~Wiedenbeck, \enquote{\textsl{The roles beacons
  play in comprehension for novice and expert programmers}}. In 14th
  \textit{Workshop Psychology of Programming Interest Group}, pp.\ 58--73, Jun
  2002.

\bibitem{curtis89}
B.~Curtis, S.~B. Sheppard, E.~Kruesi-Bailey, J.~Bailey, and D.~A. Boehm-Davis,
  \enquote{\textsl{Experimental evaluation of software documentation formats}}.
  \textit{J.\ Syst.\ \& Softw.} \textbf{9(2)}, pp.\ 167--207, Feb 1989,
  \doi{10.1016/0164-1212(89)90019-8}.

\bibitem{fluri07}
B.~Fluri, M.~{W\"ursch}, and H.~C. Gall, \enquote{\textsl{Do code and comments
  co-evolve? on the relation between source code and comment changes}}. In 14th
  \textit{Working Conf.\ Reverse Eng.}, pp.\ 70--79, Oct 2007,
  \doi{10.1109/WCRE.2007.21}.

\bibitem{head18}
A.~Head, C.~Sadowski, E.~Murphy-Hill, and A.~Knight, \enquote{\textsl{When not
  to comment: Questions and tradeoffs with {API} documentation for {C++}
  projects}}. In 40th \textit{Intl.\ Conf.\ Softw.\ Eng.}, pp.\ 643--653, May
  2018, \doi{10.1145/3180155.3180176}.

\bibitem{huang22}
Y.~Huang, M.~Wei, S.~Wang, J.~Wang, and Q.~Wang, \enquote{\textsl{Yet another
  combination of {IR}- and neural-based comment generation}}. \textit{Inf.\ \&
  Softw.\ Tech.} \textbf{152}, art.\ 107001, Dec 2022,
  \doi{10.1016/j.infsof.2022.107001}.

\bibitem{jiang06}
Z.~M. Jiang and A.~E. Hassan, \enquote{\textsl{Examining the evolution of code
  comments in {PostgreSQL}}}. In \textit{Working Conf.\ Mining Softw.\
  Repositories}, pp.\ 179--180, May 2006, \doi{10.1145/1137983.1138030}.

\bibitem{kramer99}
D.~Kramer, \enquote{\textsl{{API} documentation from source code comments: A
  case study of {JavaDoc}}}. In 17th \textit{Intl.\ Conf.\ Comput.\ Doc.}, pp.\
  147--153, Oct 1999, \doi{10.1145/318372.318577}.

\bibitem{lemos20}
O.~A.~L. Lemos, M.~Suzuki, A.~C. de~Paula, and C.~Le~Goes,
  \enquote{\textsl{Comparing identifiers and comments in engineered and
  non-engineered code: A large-scale empirical study}}. In 35th \textit{ACM
  Symp.\ Applied Computing}, pp.\ 100--109, Mar 2020,
  \doi{10.1145/3341105.3373972}.

\bibitem{lethbridge03}
T.~C. Lethbridge, J.~Singer, and A.~Forward, \enquote{\textsl{How software
  engineers use documentation: The state of the practice}}. \textit{IEEE
  Softw.} \textbf{20(6)}, pp.\ 35--39, Nov/Dec 2003,
  \doi{10.1109/MS.2003.1241364}.

\bibitem{levy21}
O.~Levy and D.~G. Feitelson, \enquote{\textsl{Understanding large-scale
  software systems –- structure and flows}}. \textit{Empirical Softw.\ Eng.}
  \textbf{26(3)}, art.~48, May 2021, \doi{10.1007/s10664-021-09938-8}.

\bibitem{liskov:book}
B.~Liskov and J.~Guttag, \textit{Abstraction and Specification in Program
  Development}. MIT Press, 1986.

\bibitem{malik08}
H.~Malik, I.~Chowdhury, H.-M. Tsou, Z.~M. Jiang, and A.~E. Hassan,
  \enquote{\textsl{Understanding the rationale for updating a function's
  comment}}. In \textit{Intl.\ Conf.\ Softw.\ Maintenance}, pp.\ 167--176, Sep
  2008, \doi{10.1109/ICSM.2008.4658065}.

\bibitem{mcconnell:cc}
S.~McConnell, \textit{Code Complete}. Microsoft Press, 2nd ed., 2004.

\bibitem{meng19}
M.~Meng, S.~Steinhardt, and A.~Schubert, \enquote{\textsl{How developers use
  {API} documentation: An observation study}}. \textit{Commun.\ Des.\ Q.}
  \textbf{7(2)}, pp.\ 40--49, Jul 2019, \doi{10.1145/3358931.3358937}.

\bibitem{meyer92}
B.~Meyer, \enquote{\textsl{Applying ``design by contract''}}. \textit{Computer}
  \textbf{25(10)}, pp.\ 40--51, Oct 1992, \doi{10.1109/2.161279}.

\bibitem{parnas94}
D.~L. Parnas, \enquote{\textsl{Software aging}}. In 16th \textit{Intl.\ Conf.\
  Softw.\ Eng.}, pp.\ 279--287, May 1994, \doi{10.1109/ICSE.1994.296790}.

\bibitem{prechelt02}
L.~Prechelt, B.~Unger-Lamprecht, M.~Philippsen, and W.~F. Tichy,
  \enquote{\textsl{Two controlled experiments assessing the usefulness of
  design pattern documentation in program maintenance}}. \textit{IEEE Trans.\
  Softw.\ Eng.} \textbf{28(6)}, pp.\ 595--606, Jun 2002,
  \doi{10.1109/TSE.2002.1010061}.

\bibitem{rai22}
S.~Rai, R.~C. Belwal, and A.~Gupta, \enquote{\textsl{A review on source code
  documentation}}. \textit{ACM Trans.\ Intelligent Syst.\ \& Tech.}
  \textbf{13(5)}, art.~84, Jun 2022, \doi{10.1145/3519312}.

\bibitem{robillard09}
M.~P. Robillard, \enquote{\textsl{What makes {API}s hard to learn? answers from
  developers}}. \textit{IEEE Softw.} \textbf{26(6)}, pp.\ 27--34, Nov/Dec 2009,
  \doi{10.1109/MS.2009.193}.

\bibitem{steidl13}
D.~Steidl, B.~Hummel, and E.~Juergens, \enquote{\textsl{Quality analysis of
  source code comments}}. In 21st \textit{Intl.\ Conf.\ Program Comprehension},
  pp.\ 83--92, May 2013, \doi{10.1109/ICPC.2013.6613836}.

\bibitem{uddin15}
G.~Uddin and M.~P. Robillard, \enquote{\textsl{How {API} documentation fails}}.
  \textit{IEEE Softw.} \textbf{32(4)}, pp.\ 68--75, Jul/Aug 2015,
  \doi{10.1109/MS.2014.80}.

\bibitem{wang23}
C.~Wang, H.~He, U.~Pal, D.~Marinov, and M.~Zhou, \enquote{\textsl{Suboptimal
  comments in {Java} projects: From independent comment changes to commenting
  practices}}. \textit{ACM Trans.\ Modeling \& Comput.\ Simulation}
  \textbf{32(2)}, art.~45, Apr 2023, \doi{10.1145/3546949}.

\bibitem{watson13}
R.~Watson, M.~Stamnes, J.~Jeannot-Schroeder, and J.~H. Spyridakis,
  \enquote{\textsl{{API} documentation and software community values: A survey
  of open-source {API} documentation}}. In 31st \textit{Intl.\ Conf.\ Design of
  Commun.}, pp.\ 165--174, Sep 2013, \doi{10.1145/2507065.2507076}.

\bibitem{woodfield81}
S.~N. Woodfield, H.~E. Dunsmore, and V.~Y. Shen, \enquote{\textsl{The effect of
  modularization and comments on program comprehension}}. In 5th \textit{Intl.\
  Conf.\ Softw.\ Eng.}, pp.\ 215--223, Mar 1981.

\end{thebibliography}

\end{document}